\documentclass{aa}
\usepackage[dvips]{graphicx}
\usepackage{psfig}
\usepackage{ltxtable}
\def\lesssim{\mathrel{\hbox{\rlap{\hbox{\lower4pt\hbox{$\sim$}}}\hbox{$<$}}}}
\def\gtrsim{\mathrel{\hbox{\rlap{\hbox{\lower4pt\hbox{$\sim$}}}\hbox{$>$}}}}
\begin{document}
\titlerunning{Importance of the ortho:para H$_2$ ratio}
\title{The importance of the ortho:para H$_2$ ratio for 
the deuteration of molecules  during pre--protostellar collapse}


\author{D.R. Flower\inst{1}
\and G. Pineau des For\^{e}ts\inst{2,3}
\and C.M. Walmsley\inst{4}}

\institute{Physics Department, The University,
           Durham DH1 3LE, UK
\and       Institut d'Astrophysique Spatiale (IAS), B\^{a}timent 121, F-91405 Orsay, France
\and       Universit\'{e} Paris-Sud 11 and CNRS (UMR 8617)
\and       INAF, Osservatorio Astrofisico di Arcetri,
           Largo Enrico Fermi 5, I-50125 Firenze, Italy}

\offprints{C.M. Walmsley}


\abstract{
 We have studied the evolution of molecular gas during the early 
stages of protostellar collapse, when the freeze--out of `heavy' 
species on to grains occurs. }
{In addition to studying the freeze--out of `heavy' species on
 to grains, we wished to compute the variation of the population 
densities of the different nuclear spin states of `tracer'
  molecular ions, such as H$_2$D$^+$ and D$_2$H$^+$, which
 are currently observed only in their ortho and para forms,
 respectively.}
{Chemical processes which determine the relative populations 
of the nuclear spin states of molecules and molecular ions 
were included explicitly.
  Nuclear spin--changing reactions have received much less
 attention in the literature than those leading to deuteration;
 but, in fact, the former processes are as significant
 as the latter and often involve the same reactants.
 A `free--fall' model of gravitational collapse 
was adopted.}
{ The ortho:para ratios of some species,
 e.g. H$_2$D$^+$, vary considerably as the density increases.
 Because the dynamical timescale is relatively short, 
there can be large departures of 
the predictions of the free--fall model from the steady--state 
solution at the same density. 
Values of the ortho:para H$_2$ ratio much higher than in steady 
state, which would prevail in `young' molecular clouds, are 
found to be inconsistent with high levels of deuteration of
 the gas. The internal energy of ortho--H$_2$ acts as a 
reservoir of chemical energy which inhibits the deuteration 
of H$_3^+$ and hence of other species,
 such as N$_2$H$^+$ and NH$_3$. }
{The principal conclusion is that the degree of deuteration 
of molecular ions and molecules 
is sensitive to the ortho:para H$_2$ ratio and hence to the
 chemical and thermal history of the precursor molecular cloud.
}

\maketitle
%

\section{Introduction}

Studies of the early stages of protostellar collapse are presently at the leading edge of research in astronomy. These studies have been driven mainly by new observations, which suggest that, during an initial, approximately isothermal phase of the collapse process, molecular species containing the heavy elements, notably carbon and oxygen, are removed from the gas phase by `freeze--out' on to the surfaces of grains (Bacmann et al. 2002, Tafalla et al. 2004). Nitrogen--bearing species, specifically N$_2$H$^+$ and NH$_3$, subsist longer (Bergin \& Langer 1997, Belloche \& Andr\'{e} 2004) but eventually freeze out also. There remains a gas which is `fully depleted', in the sense that only hydrogen--bearing species and helium remain.

In conditions under which CO is heavily depleted from the gas phase, H$^+$ and  H$_3^+$ become the principal ions. The abundance of H$^+$ is controlled by its recombination with electrons on the surfaces of negatively charged grains: two--body recombination in the gas phase is a radiative process and hence slow. Indeed, the relative abundance of H$^+$ and H$_3^+$ during collapse depends on the surface area of negatively charged grains, per unit volume of gas, which is available for the recombination of H$^+$ ions. If the rate of coagulation of the grains is insufficiently rapid, compared with the rate of collapse of the protostellar core, the grain surface area remains high and the H$^+$ ions are more effectively neutralized; H$_3^+$ becomes the dominant ion.

Observations have shown, and calculations have confirmed, that the abundances of the singly and multiply deuterated forms of H$_3^+$ can attain exceptionally high values under conditions of extreme heavy element depletion (Caselli et al. 2003, Ceccarelli et al. 2004, Vastel et al. 2004; Walmsley et al. 2004, Ceccarelli \& Dominik 2005). Furthermore, these high degrees  of deuteration can propagate to the residual nitrogen--bearing species, such as N$_2$H$^+$ and NH$_3$. The deuterated isotopes of NH$_3$ have been detected with fractional abundances which are many orders of magnitude higher than would be anticipated on the basis of the D:H elemental abundance ratio, which is of the order of $10^{-5}$. The related reaction networks have been considered in detail by several authors (Roberts et al. 2003, 2004; Roueff et al. 2005). Deuteration occurs in deuteron--proton exchange reactions with the deuterated forms of H$_3^+$. The analogous proton exchange reactions modify the relative population densities of the different possible nuclear spin states of molecules and molecular ions, such as NH$_3$ and H$_3^+$ itself. These processes have received much less attention in the literature but are, in fact, as significant as deuteration. Observations of ortho and para forms could provide information on the physical conditions in the core during collapse. Moreover, when only one of the possible nuclear spin states of a species is observed, which is usually the case, calculations of the relative population densities provide a means of estimating the total number density of the species.

The ortho:para ratio of species such as H$_3^+$ and H$_2$D$^+$ are dependent on the ortho:para H$_2$ ratio and vary with the gas density in the course of the gravitational collapse. Furthermore, the degree of deuteration of H$_3^+$ depends on the density of ortho--H$_2$, through the reaction ortho--H$_2$D$^+$(ortho--H$_2$, HD)H$_3^+$, which is the reverse of the process of deuteration of H$_3^+$ (cf. Pineau des For\^{e}ts et al. 1991, Gerlich et al. 2002). The analogous reactions involving para--H$_2$ and para--H$_2$D$^+$ are endothermic and slow at low gas temperatures. However, para--H$_2$D$^+$ is converted into ortho--H$_2$D$^+$ in the reaction para--H$_2$D$^+$(ortho--H$_2$, para--H$_2$)ortho--H$_2$D$^+$, which is exothermic. Thus, as the assumed value of the initial abundance of ortho--H$_2$ (equivalently, the initial ortho:para H$_2$ ratio) increases, both ortho-- and para--H$_2$D$^+$ are converted back to H$_3^+$. Consequently, the initial degrees of deuteration, not only of H$_3^+$, but also of nitrogen--bearing species such as N$_2$H$^+$ and NH$_3$, fall to low values, as H$_2$D$^+$ is the main agent of their deuteration. It should be clear from these comments that  the initial value of the ortho:para H$_2$ ratio is critical to the composition of the collapsing gas, particularly to its level of deuteration. Indeed, the high levels of deuteration which are observed in some protostellar cores impose constraints on the initial ortho:para H$_2$ ratio.

Molecular hydrogen forms on grains, at a rate which is still not well determined. It is believed (and assumed here) that the ortho:para ratio of the hydrogen molecules which are formed and released into the gas phase is the statistical value of 3:1. Subsequent proton--exchange reactions with H$^+$ and H$_3^+$ interconvert ortho-- and para--H$_2$. However, at low temperatures, the timescales associated with these reactions are large, owing to the low degree of ionization of the gas; this is particularly true of para to ortho conversion, which is endothermic. As we shall see below, it is certainly {\it not} clear that the ortho:para H$_2$ ratio at the commencement of gravitational collapse of the molecular gas has attained its value in steady state; and it is almost certainly higher than its value in thermal (Boltzmann) equilibrium. At low temperatures, ortho--H$_2$ is formed principally by H--atom recombination on the surfaces of grains, rather than through proton--exchange reactions with para--H$_2$ in the gas phase. Accordingly, we have treated the initial ortho:para H$_2$ ratio as a parameter, which we have varied between its upper limit of 3 and its lower limit, taken to be the ratio of the populations of the $J = 1$ and $J = 0$ rotational states in thermal equilibrium at kinetic temperature $T$. A high fractional abundance of ortho--H$_2$ constitutes a major source of (internal) energy in molecular gas at low kinetic temperatures ($T \approx 10$~K): the lowest state of ortho--H$_2$, $J = 1$, lies approximately 170~K above the ground state of para--H$_2$, $J = 0$; this energy becomes available in proton--exchange reactions with ortho--H$_2$.

In Section 2, we outline the model of gravitational collapse which we have employed and consider the processes which influence the ortho:para ratios (more generally, the relative populations of the possible nuclear spin states of a given molecule or molecular ion). Section 3 contains the results of the calculations of the fractional abundances of the ions and of the relative populations of the nuclear spin states. Particular attention is paid to the influence of the initial value of the ortho:para H$_2$ ratio. Also, we compare the results of the calculations, following the gravitational collapse, with the values which are predicted, assuming that the medium has time to attain a steady state. The possible effects of grain coagulation are considered. In Section 4, we summarize our results and make our concluding remarks.

\section{The model}
\label{Model}

Following the procedures and conclusions of our previous paper (Flower et al. 2005), we consider the evolution of a contracting sphere of gas and dust and apply the equations appropriate to free--fall, homologous, isothermal collapse. The timescale for free--fall collapse is 

\begin{equation}
\tau _{\rm ff} = \left [ \frac {3\pi }{32G\rho _0} \right ]^{\frac {1}{2}}
\label{equ1}
\end{equation}
where $\rho _0$ is the initial mass per unit volume. For an initial gas density of $n_{\rm H} = n({\rm H}) + 2n({\rm H}_2) = 10^4$~cm$^{-3}$, the free--fall time is of the order of $10^5$~yr. The (constant) temperature was $T = 10$~K. These values of $n_{\rm H}$ and $T$ are typical of those prevailing in dark molecular clouds, from which protostars form, through gravitational contraction.

The initial composition was taken to be that of a molecular gas in static equilibrium, with the elemental abundances and depletions as specified in table 1 of Flower et al. (2005). The value of the dust:gas mass ratio was 0.0094. The chemical reaction set was a composite of our previous studies (Walmsley et al. 2004, Flower et al. 2004, 2005). It comprises reactions involving species containing H, He, C, N, O and S and distinguishes the nuclear spin states of H$_3^+$ and its deuterated isotopes, in addition to those of H$_2^+$ and H$_2$ (see Section~\ref{statistics} below). Following the discussion in Section 4.4 of Flower et al. (2005), we have continued to adopt a grain sticking coefficient $S({\rm N}) = 0.1$ for atomic nitrogen, and, with a view to internal consistency, we have taken $S({\rm O}) = 0.1$ also. For all other atomic and molecular species, $S = 1.0$. The rate of cosmic ray ionization of hydrogen was taken to be $\zeta = 1\times 10^{-17}$ s$^{-1}$, as in our previous paper.

The initial grain radius was taken to be $a_{\rm g} = 0.05$ $\mu $m. As shown in our previous paper (Flower et al. 2005), this value of $a_{\rm g}$ yields approximately the same grain opacity, $n_{\rm g}\pi a_{\rm g}^2$ (where $n_{\rm g}$ is the grain number density), as the grain size distribution of Mathis et al. (1977), with limits of $0.01 \le a_{\rm g} \le 0.3$ $\mu $m.

The validity of assuming static equilibrium, particularly for the ortho:para H$_2$ ratio, will be considered below. In view of the importance of the ortho:para H$_2$ ratio -- which becomes apparent from the results presented in Section~\ref{Results} -- we have varied its initial value between the statistical ratio of 3 and the ratio of $3.6\times 10^{-7}$, which corresponds to Boltzmann equilibrium [Equ.~(\ref{equ2}) below] at $T = 10$~K.

\subsection{Nuclear spin statistics}
\label{statistics}

In the calculations reported below, the abundances not only of the chemical species but also of their individual nuclear spin states have been determined, where applicable. The cases of interest here involve molecules or molecular ions comprising two or more protons, such as H$_2$. Fermi--Dirac statistics require that the nuclear wave function should be asymmetric under exchange of identical protons, and this restriction associates states of given nuclear spin symmetry with states of appropriate rotational symmetry. Thus, in the case of H$_2$, states with total nuclear spin $I = 0$ (para--H$_2$) are associated with rotational states with even values of the rotational quantum number, $J$; states with $I = 1$ (ortho--H$_2$) are associated with odd $J$. Transitions between states of differing nuclear spin, $I = 0$ and $I = 1$ in this example, are induced by proton--exchange reactions with H$^+$ and H$_3^+$. [Although HCO$^+$ is initially the most abundant ion which might undergo proton exchange with H$_2$, the measurement by Huntress (1977) of the rate coefficient at $T = 300$~K for the reaction HCO$^+$(D$_2$, HD)DCO$^+$ suggests that the analogous reaction with H$_2$ is probably too slow to be significant.] In view of the importance of H$_2$ in the chemistry of molecular clouds, it might be anticipated that its ortho:para ratio would influence the analogous ratios in other molecules and molecular ions.

In order to provide a framework for the interpretation of the numerical results which will be presented in the following Section, we shall establish first the values of the relative populations of the nuclear spin states of a number of key species, in static equilibrium, on the basis of an analysis which is necessarily statistical in nature.

\subsubsection{H$_2$}

We assume that ortho-- and para--H$_2$ are formed (on grains) in the statistical ratio of 3:1, i.e. the ratio of the corresponding values of $(2I + 1)$. Subsequent radiative cascade leads to the $J = 1$ ($I = 1$) and $J = 0$ ($I = 0$) levels of the ground vibrational state, which are separated by 170.5~K. The rates of cosmic ray dissociation and ionization of H$_2$ are taken to be independent of the total nuclear spin, $I$. At the low temperature which we consider ($T = 10$~K), proton--exchange reactions convert ortho--H$_2$ to para--H$_2$; the reverse reaction is insufficiently rapid at low $T$ for the ortho:para ratio to attain its Boltzmann value,

\begin{equation}
\frac {n(J=1)}{n(J=0)} = 9\ {\rm exp} \left (\frac {-170.5}{T} \right )
\label{equ2}
\end{equation}
where the factor of 9 corresponds to the ratio of the statistical weights, $(2I + 1)(2J + 1)$, of these ortho ($J = 1$) and para ($J = 0$) levels. At $n_{\rm H} = 10^4$~cm$^{-3}$, the steady--state value of $n(J = 1)/n(J = 0)$ is $2.7\times 10^{-3}$, which is much larger than the value of this ratio ($3.6\times 10^{-7}$) in Boltzmann equilibrium at $T = 10$~K. In other words, at low temperatures, ortho--H$_2$ is formed principally by H--atom recombination on the surfaces of grains, rather than from para--H$_2$ in the gas phase, through proton--exchange reactions with H$^+$ and H$_3^+$. At a higher temperature of $T = 30$~K, the steady--state value of $n(J = 1)/n(J = 0)$ ($3.3\times 10^{-2}$) is much closer to its value in Boltzmann equilibrium at this temperature ($3.1\times 10^{-2}$; cf. Fig.~\ref{s-s1}), as would be expected.

\subsubsection{H$_2^+$}

H$_2^+$ is formed through cosmic ray ionization of H$_2$, and hence the rates of production of ortho-- and para--H$_2^+$ are proportional to the number densities of ortho-- and para--H$_2$, respectively. Both ortho-- and para--H$_2^+$ are destroyed by para-H$_2$, which is the much more abundant of the two forms of H$_2$, at equal rates, yielding H$_3^+$. It follows that the ortho:para ratio of H$_2^+$ is essentially the same as that of H$_2$.

\subsubsection{H$_3^+$}

Para--H$_3^+$ forms principally from para--H$_2^+$ and para--H$_2$, which are the more abundant forms of H$_2^+$ and H$_2$, respectively, 

\begin{equation}
{\rm H}_2^+({\rm p}) + {\rm H}_2({\rm p}) \rightarrow {\rm H}_3^+({\rm p}) + {\rm H}
\label{equ3}
\end{equation}
whereas ortho--H$_3^+$ is produced from either ortho--H$_2^+$ and para--H$_2$ or para--H$_2^+$ and ortho--H$_2$, 

\begin{equation}
{\rm H}_2^+({\rm o}) + {\rm H}_2({\rm p}) \rightarrow {\rm H}_3^+({\rm o}) + {\rm H}
\label{equ4}
\end{equation}
\begin{equation}
{\rm H}_2^+({\rm p}) + {\rm H}_2({\rm o}) \rightarrow {\rm H}_3^+({\rm o}) + {\rm H}
\label{equ5}
\end{equation}
with the same total rate coefficient ($2.1\times 10^{-9}$~cm$^3$ s$^{-1}$). Recalling that the ortho:para ratio of H$_2^+$ is essentially the same as that of H$_2$, we conclude that the ortho:para H$_3^+$ ratio expected from considerations of its formation and destruction alone is the same as the ortho:para H$_2$ ratio. However, proton--exchange reactions of H$_3^+$ with H$_2$ modify the ortho:para H$_3^+$ ratio. The proton--exchange reactions of the lowest energy states of ortho-- and para--H$_3^+$ with para--H$_2$ are both endoergic, by 137.6~K and 203.4~K, respectively, and consequently negligible at $T = 10$~K. On the other hand, the reverse reactions with ortho--H$_2$ 

\begin{equation}
{\rm H}_3^+({\rm o}) + {\rm H}_2({\rm o}) \rightarrow {\rm H}_3^+({\rm p}) + {\rm H}_2({\rm p})
\label{equ6}
\end{equation}
\begin{equation}
{\rm H}_3({\rm p})^+ + {\rm H}_2({\rm o}) \rightarrow {\rm H}_3^+({\rm o}) + {\rm H}_2({\rm p})
\label{equ7}
\end{equation}
are exoergic and would yield a ratio of ortho:para H$_3^+$ of 4:1, as may be seen from the following argument. 

The statistical weight of the lowest level of para--H$_3^+$ is 2, and that of ortho--H$_3^+$ is 4. The rate coefficient for a reaction is taken proportional to the ratio of the combined (by multiplication) statistical weight of the products to that of the reactants. Thus, the rate coefficient for the reaction (\ref{equ7}) is taken to be 4 times larger than for reaction (\ref{equ6}). 

In equilibrium, the population densities of ortho-- and para--H$_3^+$ are found to be comparable, i.e. the proton--exchange reactions with ortho--H$_2$ dominate. As a consequence, the ortho:para H$_3^+$ ratio depends on the density of ortho--H$_2$ and hence on ortho:para H$_2$ ratio.

\section{Results}
\label{Results}

\subsection{Initial abundances}
\label{Tcr}

The elemental abundances and their initial repartition between the gas and the solid phases were taken from table 1 of Flower et al. (2005); the value of the dust:gas mass ratio was 0.0094. The chemical composition of the gas, in steady state, was determined by running a time--dependent calculation, at a constant density $n_{\rm H} = 10^4$~cm$^{-3}$ and kinetic temperature $T = 10$~K, neglecting any further freeze--out on to the grains, until the fractional abundances became constant. Under these conditions, the steady--state value of the ortho:para H$_2$ ratio is $2.7\times 10^{-3}$.

Flower \& Pineau des For\^{e}ts (1990) and Pineau des For\^{e}ts et al. (1991) studied the isobaric thermal and chemical evolution of interstellar clouds, from initially atomic to finally molecular gas which had attained a steady state; the values of the gas pressure which they considered were similar to the initial value  adopted here ($n_{\rm H}T = 10^5$~cm$^{-3}$~K). Their studies showed that the timescale for establishing the steady--state value of the ortho:para H$_2$ ratio is particularly large, in excess of $10^7$~yr. At time $t \approx 10^6$~yr, the formation of H$_2$ on grains gives rise to a local maximum of the ortho:para ratio, which approaches the statistical value of 3:1 associated with the formation process. The ortho:para ratio subsequently falls towards its steady--state value, which is reached at $t \approx 10^7$~yr; the steady--state value of the ratio exceeds that in Boltzmann equilibrium [Equ.~(\ref{equ2})]. Thus, during the time interval $10^6 \lesssim t \lesssim 10^7$~yr, the ortho:para H$_2$ ratio in the molecular gas remains larger than its final, steady--state value.

\begin{figure}
\centering
\includegraphics[height=12cm]{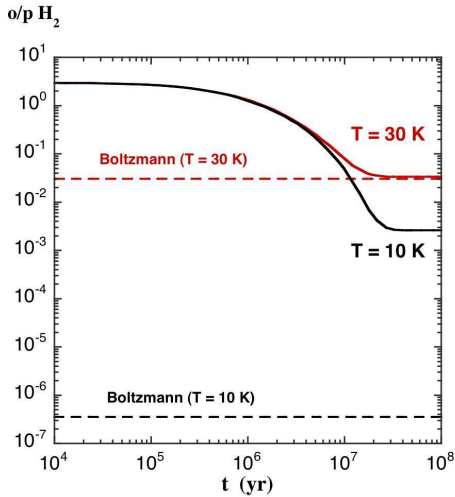}
\caption{The ratio of abundances of ortho-- and para--H$_2$ as a function of time, $t$, from an assumed initial value of 3 through to the attainment of steady state. A constant gas density of $n_{\rm H} = 10^4$~cm$^{-3}$ and temperatures $T = 10$~K and $T = 30$~K were adopted. The ortho:para ratios in Boltzmann equilibrium are shown for both values of $T$.}
\label{s-s1}
\end{figure}

In Fig.~\ref{s-s1} is plotted the ratio of ortho-- to para--H$_2$ as a function of time, $t$, at a constant gas density of $n_{\rm H} = 10^4$~cm$^{-3}$ and kinetic temperatures $T = 10$~K and $T = 30$~K; the initial value was taken to be 3, corresponding to the statistical ratio associated with the formation of H$_2$ on grains. Fig.~\ref{s-s1} shows that the time required for the ratio of ortho-- to para--H$_2$ to attain its steady--state value ($2.7\times 10^{-3}$ at $T = 10$~K and $3.3\times 10^{-2}$ at $T = 30$~K) is approximately $3\times 10^7$~yr; this time is comparable with a recent determination, by Tassis \& Mouschovias (2004), of the lifetimes of molecular clouds ($\approx 10^7$~yr). [We note that Hartmann et al. (2001) estimated the ages of clouds in the solar vicinity as being of the order of 10$^6$ yr.] We conclude that it is probable that the ortho:para H$_2$ ratio does not reach its steady--state value before protostellar collapse begins. Accordingly, we have varied the initial value of the ortho:para H$_2$ ratio between the statistical value of 3 and $3.6\times 10^{-7}$, corresponding to Boltzmann equilibrium at $T = 10$~K. It transpires that the initial value of this ratio is critical not only to the populations of the nuclear spin states of H$_3^+$ and its deuterated isotopes but, perhaps more significantly, also to the degree of deuteration of H$_3^+$ (cf. Pineau des For\^{e}ts et al. 1991).

We see from Fig.~\ref{s-s1} that, for $T = 30$~K, the ortho:para H$_2$ ratio in steady--state is close to its value in thermodynamic equilibrium, whereas this is not the case for $T = 10$~K. At low temperatures, the direct formation of ortho--H$_2$ on grains is more rapid than the endoergic proton--exchange reactions of H$^+$ and H$_3^+$ with para--H$_2$. The value of the kinetic temperature, $T_{\rm cr}$, below which the steady--state value of the ortho:para ratio exceeds its value in Boltzmann equilibrium may be estimated by equating these two rates of formation of ortho--H$_2$. Noting that ortho-- and para--H$_2$ are assumed to form, on grains, in the statistical ratio of 3:1, and equating the total rate of formation of H$_2$ to the rate of its destruction by cosmic rays, we obtain $T_{\rm cr} \approx 20$~K for the parameters adopted in the present calculations. In general, when $T \lesssim T_{\rm cr}$, the initial ortho:para H$_2$ ratio depends on the thermal history of the core.

\subsection{Depletion during collapse}

\begin{figure}
\centering
\includegraphics[height=12cm]{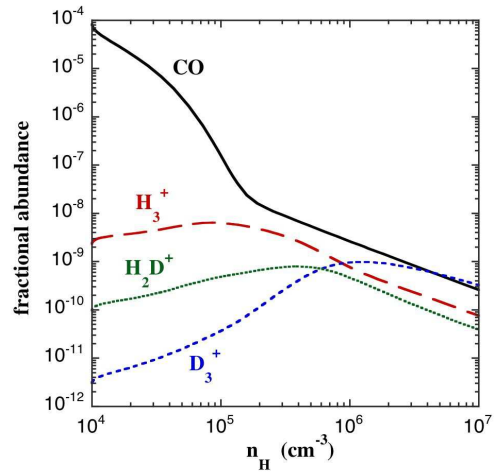}
\caption{The fractional abundances of CO and selected isotopes of H$_3^+$ as functions of the gas density, $n_{\rm H}$, for a constant kinetic temperature, $T = 10$~K (our `reference' model of collapse). The initial value of the ortho:para H$_2$ ratio is $2.7\times 10^{-3}$.}
\label{rm1}
\end{figure}

In Fig.~\ref{rm1}, we show the fractional abundances of CO and selected isotopes of H$_3^+$ as functions of the gas density, $n_{\rm H}$, for our `reference' model of collapse. We note that density is related to age through the equation for free--fall collapse. 

The tracer molecule, CO, is already significantly depleted (through freeze--out) by the time that $n_{\rm H} = 3\times 10^4$ cm$^{-3}$, whereas the fractional abundances of H$_2$D$^+$ and D$_3^+$ increase until much higher densities, $n_{\rm H} > 3\times 10^5$ cm$^{-3}$, corresponding to later times. It is clear that, in this model, deuterium fractionation occurs at densities for which CO is practically completely depleted (as is N$_2$ also). In order to achieve high levels of deuterium fractionation in molecules such as NH$_3$, it is necessary to minimize the difference between the critical densities (or the corresponding times) at which (i) heavy species, such as CO and N$_2$, have frozen out, and (ii) deuterium fractionation, through reactions with H$_2$D$^+$ and the other deuterated forms of H$_3^+$, has begun. 

\subsection{Steady--state vs. free--fall}

In our previous paper (Flower et al. 2005, fig. 6), we compared the fractional abundances of H$^+$, H$_3^+$ and e$^-$, in steady--state and as predicted by the free--fall model, for densities $2\times 10^5 \le n_{\rm H} \le 2\times 10^7$~cm$^{-3}$. The comparison was made in the limit of complete heavy element depletion. At times which are sufficiently large for steady state to be attained, essentially all the species containing elements heavier than He are frozen on to the grains. The agreement between the two calculations was found to be good, for the species listed above; but the deuterated isotopes of H$_3^+$ were not included in the collapse model, and neither were the various possible nuclear spin modifications. We are now in a position to make such comparisons in considerably more detail. However, for reasons which will now be given, we have chosen to make the comparison in a somewhat different way from that followed in our previous paper.

The free--fall collapse is initiated at a density $n_{\rm H} = 10^4$~cm$^{-3}$ and for a given repartition of the elements between the gas and solid phases (cf. Section 3.1 above). In the course of the free--fall collapse, atoms and molecules continue to freeze on to the grains, on a timescale which is comparable with the free--fall time. Thus, the degree of depletion of the heavy elements from the gas phase increases with time and hence with the density of the medium. When comparing the `collapse' model with the `steady--state' solution, we adopt the composition of the gas at a given density on the collapse profile and then compute the corresponding steady--state solution, at a time $t > 10^8$~yr, keeping constant both the density and the degree of depletion of the gas. This procedure ensures that the comparison reflects departures of the free--fall model from steady state for {\it all} times (densities), including those which fall short of the `complete depletion' limit.

\subsubsection{Fractional abundances of ions}

\begin{figure}
\centering
\includegraphics[height=12cm]{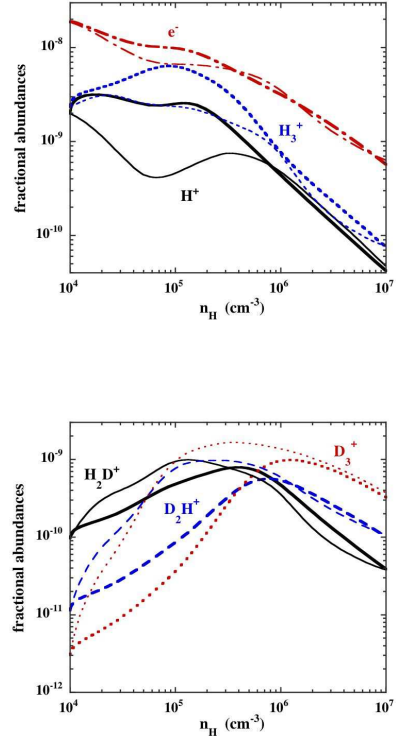}
\caption{The fractional abundances, relative to $n_{\rm H}$, of (a) H$^+$ (full curve), H$_3^+$ (broken curve) and e$^-$ (chain curve), and (b) H$_2$D$^+$ (full curve), D$_2$H$^+$ (long--dashed curve) and D$_3^+$ (short--dashed curve), as functions of $n_{\rm H}$. The bold curves are the predictions of the free--fall model, the light curves are the corresponding results in steady--state.}
\label{s-s2}
\end{figure}

First, we compare the abundances of H$^+$ and H$_3^+$, computed in steady--state and following the free--fall collapse: see Fig.~\ref{s-s2}a. In the present calculations, the deuterated isotopes of H$_3^+$ and their nuclear spin modifications have been included in the calculations. For the purposes of this first comparison, we sum the fractional abundances of ortho-- and para--H$_3^+$. Reactions leading to the deuteration of heavy species (those containing elements heavier than He, specifically NH$_3$ and N$_2$H$^+$) have not been included explicitly in the present model. Nonetheless, qualitative conclusions regarding the degrees of deuteration of such species can still be drawn, in so far as the deuterated isotopes of H$_3^+$ are primarily responsible for the deuterium fractionation of heavy molecules. We note that the abundances of the deuterated isotopes of H$_3^+$ are determined mainly by reactions such as H$_3^+$(HD, H$_2$)H$_2$D$^+$, which determine the degree of deuteration of H$_3^+$, and not by reactions with heavy molecules, such as CO, whose fractional abundances become rapidly much smaller than that of HD, owing to freeze--out. Comparison calculations have shown that, at densities $n_{\rm H} \gtrsim 10^5$~cm$^{-3}$, at which the fractional abundances of the deuterated isotopes of H$_3^+$ become significant (cf. Fig.~\ref{rm1}), the reactions with CO have no effect on their fractional abundances.

The results of the two calculations (steady--state and free--fall collapse) are in agreement for densities $n_{\rm H} \gtrsim 10^6$~cm$^{-3}$, for which complete freeze--out has effectively occurred.  At lower densities, $n_{\rm H} \approx 10^5$~cm$^{-3}$, the steady--state solutions underestimate the fractional abundances of both H$^+$ and H$_3^+$ -- the former by approximately a factor of 5. Protons are removed by charge transfer reactions with neutral species whose ionization potentials are less than that of atomic hydrogen. As the heavy neutrals freeze on to the grains, the fractional abundance of H$^+$ begins to increase, and ultimately the steady--state solution merges with the results of the collapse model. The timescale for establishing ionization equilibrium is small (of the order of 1000 yr, for $n_{\rm H} = 10^4$~cm$^{-3}$) compared with the equilibrium timescale of the chemistry as a whole (Walmsley et al. 2004). Consequently, the free electron density calculated by the collapse model exceeds the steady--state value by no more than about 50\% over the entire range of density, $n_{\rm H}$. 

The reason that the steady--state solution underestimates the fractional abundance of H$_3^+$ in the vicinity of $n_{\rm H} = 10^5$~cm$^{-3}$ may be seen by referring to Fig.~\ref{s-s2}b, where the fractional abundances of the deuterated isotopes of H$_3^+$ are plotted. In essence, the steady--state solutions overestimate the degree of deuteration of H$_3^+$ at intermediate densities. The level of agreement between the steady--state and free--fall collapse calculations deteriorates as the degree of deuteration of H$_3^+$ increases, i.e. in the sense H$_2$D$^+$, D$_2$H$^+$, D$_3^+$; see Fig.~\ref{s-s2}b. The differences between the steady--state abundances of these species, and those calculated following the free--fall collapse, are significant in the context of the deuteration of molecules such as NH$_3$, for which deuteron--transfer and proton--deuteron exchange reactions with the deuterated isotopes of H$_3^+$ are crucially important. We note that the analogous proton--transfer and proton--exchange reactions with H$_3^+$, H$_2$D$^+$ and D$_2$H$^+$ contribute to ortho--para conversion in NH$_3$. It follows that the issues of the degree of deuteration of NH$_3$, on the one hand, and the relative population densities of its ortho and para nuclear spin states, on the other hand, are inter-related (cf. Pineau des For\^{e}ts et al. 1991).

\subsubsection{Relative population densities of nuclear spin states}

\begin{figure}
\centering
\includegraphics[height=12cm]{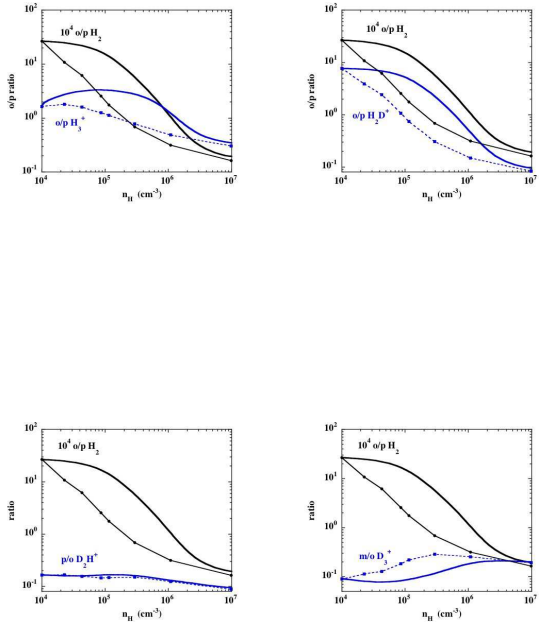}
\caption{The relative abundances of the two lowest nuclear spin states (in the sense of energetically higher:lower) of H$_3^+$, H$_2$D$^+$, D$_2$H$^+$ and D$_3^+$ as functions of $n_{\rm H}$, as predicted by the free--fall model (bold curves). The corresponding values in steady--state are shown as the light curves joining the symbols (full circles for H$_2$, full squares for the other species).}
\label{s-s3}
\end{figure}

In Fig.~\ref{s-s3} are compared the relative populations of the two lowest nuclear spin states of H$_3^+$, H$_2$D$^+$, D$_2$H$^+$ and D$_3^+$. In each case, the ortho:para H$_2$ ratio [specifically, the ratio $n(J = 1)$ : $n(J = 0)$] is plotted, for reference. We see from Fig.~\ref{s-s3} that there are large discrepancies between the predictions of the free--fall collapse model and the equivalent steady--state calculations. The ortho:para H$_3^+$ and H$_2$D$^+$ ratios are correlated with that of H$_2$, owing to the importance of proton exchange with H$_2$. Thus, the underestimation (in steady--state) of the ortho:para H$_2$ ratio results in the ortho:para H$_3^+$ and H$_2$D$^+$ ratios being underestimated also -- by more than an order of magnitude in the case of H$_2$D$^+$ at $n_{\rm H} = 10^5$~cm$^{-3}$. [We note that the ortho:para H$_2$D$^+$ ratio is initially large (at $n_{\rm H} = 10^4$~cm$^{-3}$), owing to proton--exchange reactions with ortho--H$_2$, which are statistically favourable to the production of ortho--H$_2$D$^+$ (Walmsley et al. 2004).]

In steady state, the relative value of the population densities of the meta and ortho nuclear spin states of D$_3^+$ is determined principally by deuteron exchange reactions with HD, which favour energetically the production of ortho--D$_3^+$ (specifically, the ground state of the ion; the first excited state is the lowest state of meta symmetry). However, as noted in Appendix B of Flower et al. (2004), the reaction meta--D$_3^+$(ortho--H$_2$, HD)ortho--D$_2$H$^+$, which is endoergic by only 18~K, removes meta--D$_3^+$, even at low $T$. Whilst this reaction is of secondary importance {\it in steady state} at $n_{\rm H} = 10^5$~cm$^{-3}$, owing to the decrease in the ortho:para H$_2$ ratio with increasing $n_{\rm H}$, it dominates the removal of meta--D$_3^+$ in the free--fall model at $n_{\rm H} = 10^5$~cm$^{-3}$, for which the ortho:para H$_2$ ratio is much higher than in steady state (see Fig.~\ref{s-s3}). As a consequence, the meta:ortho D$_3^+$ ratio is lower at $n_{\rm H} = 10^5$~cm$^{-3}$ in the free--fall model than in steady state at the same density.

\subsection{Dependence on the initial ortho:para H$_2$ ratio}

The ortho:para H$_2$ ratio is important in establishing the relative populations of the nuclear spin states of a number of species, as the discussion in Section~\ref{Model} should have made clear. Furthermore, the initial value of this ratio proves to be critical to the degree of deuteration of H$_3^+$ during free--fall collapse. For the purposes of the illustrations which follow, grain coagulation has been neglected.

\begin{figure}
\centering
\includegraphics[height=12cm]{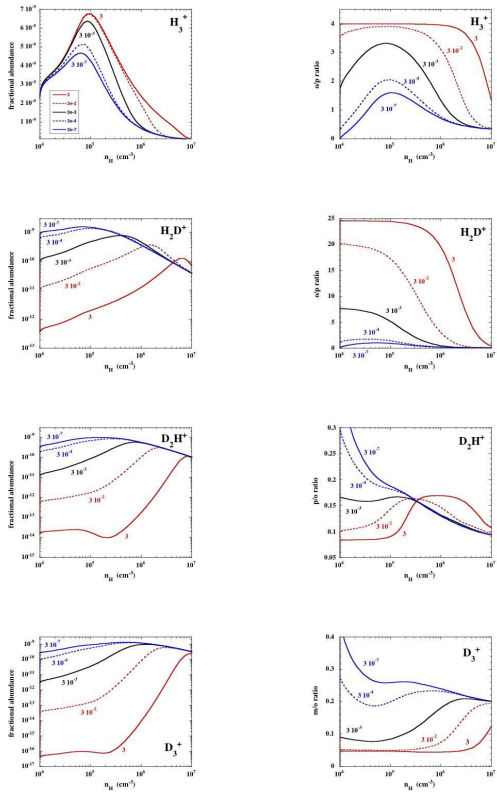}
\caption{The fractional abundances, relative to $n_{\rm H} \approx n({\rm H}) + 2n({\rm H}_2)$, of H$_3^+$, H$_2$D$^+$, D$_2$H$^+$ and D$_3^+$ as functions of the gas density, $n_{\rm H}$, at a constant temperature $T = 10$~K. Results are shown for five values of the ortho:para H$_2$ ratio, between the statistical value of 3 and $3.6\times 10^{-7}$, the ratio in Boltzmann equilibrium at $T = 10$~K. The computed values of the relative populations of nuclear spin states of the same species are also shown, in the sense of the population of the energetically higher state divided by that of the lower.}
\label{op1}
\end{figure}

In Fig.~\ref{op1} are shown both the fractional abundances and the relative populations of the nuclear spin states of H$_3^+$, H$_2$D$^+$, D$_2$H$^+$ and D$_3^+$. It is apparent from this Figure that the fractional abundances of H$_2$D$^+$ and more highly deuterated isotopes of H$_3^+$ are strongly dependent on the initial ortho:para H$_2$ ratio -- whose value is varied between 3 and $3.6\times 10^{-7}$ (the ratio in Boltzmann equilibrium at $T = 10$~K) -- even at densities as high as $n_{\rm H} = 10^6$~cm$^{-3}$. This dependence arises from the reactions

\begin{equation}
{\rm H}_2{\rm D}^+({\rm o}) + {\rm H}_2({\rm o}) \rightarrow {\rm H}_3^+({\rm p}) + {\rm HD}
\label{equ17}
\end{equation}
and

\begin{equation}
{\rm H}_2{\rm D}^+({\rm o}) + {\rm H}_2({\rm o}) \rightarrow {\rm H}_3^+({\rm o}) + {\rm HD}
\label{equ18}
\end{equation}
which remove directly ortho--H$_2$D$^+$ and indirectly para--H$_2$D$^+$, which is converted into ortho--H$_2$D$^+$ by the proton--exchange reaction 

\begin{equation}
{\rm H}_2{\rm D}^+({\rm p}) + {\rm H}_2({\rm o}) \rightarrow {\rm H}_2{\rm D}^+({\rm o}) + {\rm H}_2({\rm p})
\label{equ19}
\end{equation}
which occurs, once again, with ortho--H$_2$. Reactions~(\ref{equ17}) and (\ref{equ18}) are the reverse of the reactions of ortho-- and para--H$_3^+$ with HD, which are responsible for the deuteration of H$_3^+$; as the initial ortho:para H$_2$ ratio increases, the deuteration of H$_3^+$ is inhibited effectively by these reverse reactions. 

From Fig.~\ref{op1}, we see that the populations of the energetically higher nuclear spin states (i.e. the ortho states) of H$_3^+$ and H$_2$D$^+$ increase, relative to the lower (para) states, as the ortho:para H$_2$ ratio rises, whereas it is the {\it lower} states of the mutiply--deuterated isotopes, D$_2$H$^+$ and D$_3^+$, which become relatively more populated. Proton--exchange reactions, such as (\ref{equ19}) and the analogous reaction involving H$_3^+$, are enabled by the internal excitation energy (170~K) of ortho--H$_2$, which acts as a reservoir of chemical energy. The net effect is the transfer of population to the excited (ortho) states of H$_2$D$^+$ and H$_3^+$. In the case of H$_2$D$^+$, essentially all of the population is in the ortho state at low gas densities and high ortho:para H$_2$ ratios. On the other hand, the excited nuclear spin states of the multiply--deuterated species, D$_2$H$^+$ and D$_3^+$, are {\it removed} preferentially by ortho--H$_2$ in the reactions

\begin{equation}
{\rm D}_2{\rm H}^+({\rm p}) + {\rm H}_2({\rm o}) \rightarrow {\rm H}_2{\rm D}^+({\rm p}) + {\rm HD}
\label{equ20}
\end{equation}

\begin{equation}
{\rm D}_3^+({\rm m}) + {\rm H}_2({\rm o}) \rightarrow {\rm D}_2{\rm H}^+({\rm o}) + {\rm HD}
\label{equ21}
\end{equation}
Reaction~(\ref{equ20}) is exothermic, and reaction~(\ref{equ21}) is endothermic, but by only 18~K. (The corresponding reactions of ortho--H$_2$ with ortho--D$_2$H$^+$ and ortho--D$_3^+$ are less energetically favourable, as the reacting ion is in its ground state.)

\begin{figure}
\centering
\includegraphics[height=12cm]{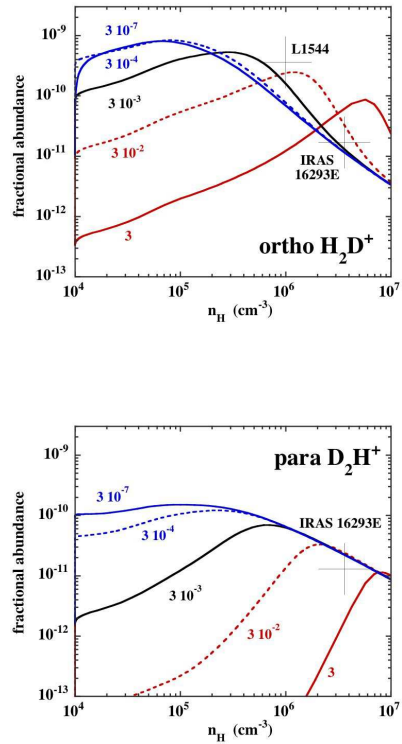}
\caption{The fractional abundances, relative to $n_{\rm H} \approx n({\rm H}) + 2n({\rm H}_2)$, of ortho--H$_2$D$^+$ and para--D$_2$H$^+$ as functions of the gas density, $n_{\rm H}$, at a constant temperature $T = 10$~K. Results are shown for five initial (i.e. at $n_{\rm H} = 10^4$ cm$^{-3}$) values of the ortho:para H$_2$ ratio, between the statistical value of 3 and $3.6\times 10^{-7}$, the ratio in Boltzmann equilibrium at $T = 10$~K. The fractional abundance of ortho--H$_2$D$^+$ in L1544 (Caselli et al. 2003) and IRAS 16293E (Vastel et al. 2004), and the fractional abundance of para--D$_2$H$^+$ in IRAS 16293E (Vastel et al. 2004), are indicated, as are the densities believed to prevail in these cores.}
\label{op2}
\end{figure}

In Fig.~\ref{op2} are plotted the fractional abundances of ortho--H$_2$D$^+$ and para--D$_2$H$^+$ as functions of the gas density, $n_{\rm H}$, in the course of the collapse. Both these species have been observed in protostellar cores, with fractional abundances of the order of $10^{-10}$ (Caselli et al. 2003, Vastel et al. 2004); the fractional abundances of ortho--H$_2$D$^+$ in L1544 (Caselli et al. 2003) and IRAS 16293E (Vastel et al. 2004), and the fractional abundance of para--D$_2$H$^+$ in IRAS 16293E, deduced from the observations of Vastel et al. (2004), are indicated. The ortho:para H$_2$ ratio which is inferred from the observations is sensitive to the gas density. In the case of IRAS 16293E, if one adopts a density of $4\times 10^6$ cm$^{-3}$ (cf. Lis et al. 2002), all models with an initial ortho:para H$_2$ ratio below 3 are consistent with the observations. For the lower density of L1544 ($6\times 10^5 \le n_{\rm H} \le 2\times 10^6$ cm$^{-3}$), the data suggest an initial ortho:para ratio which is within an order of magnitude of its steady--state value of $2.7\times 10^{-3}$ at $n_{\rm H} = 10^4$ cm$^{-3}$. More definitive conclusions must await not only further observations but also the introduction of temperature and density gradients into the models of the  cores.

\subsection{Influence of coagulation}
\label{coag}

\begin{figure}
\centering
\includegraphics[height=12cm]{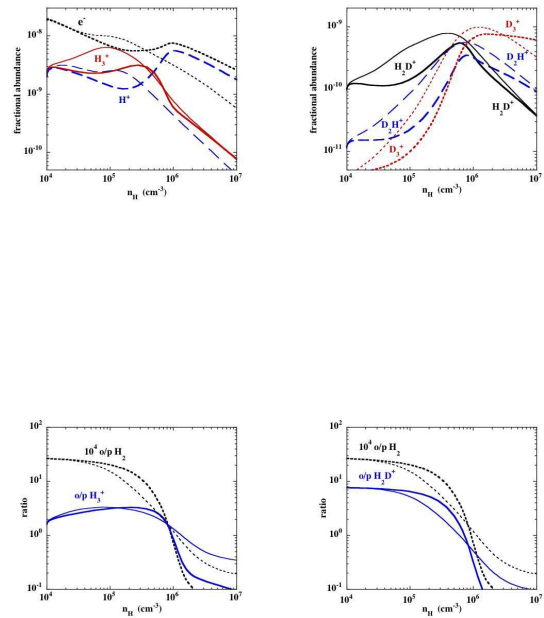}
\caption{The fractional abundances of (a) H$_3^+$ and (b) its isotopes, as functions of $n_{\rm H}$, as predicted by the free--fall model including (bold curves) and excluding (corresponding light curves) coagulation; see Section~\ref{coag}. The fractional abundances of H$^+$ and of the free electrons are also plotted in panel (a). The ortho:para ratios of H$_2$, H$_3^+$ and H$_2$D$^+$ are displayed in panels (c) and (d).} 
\label{coagulation}
\end{figure}

The coagulation of grains in the course of the collapse was considered in our previous paper (Flower et al. 2005). The critical velocity, below which two colliding grains of radius $a_{\rm g}$ are assumed to coagulate, was taken to be 

\begin{displaymath}
v_{\rm crit} \propto a_{\rm g}^{-\frac {5}{6}}
\end{displaymath}
following Chokshi et al. (1993). In the extreme cases which we considered, the constant of proportionality was taken equal to either 0 or 0.4 (with $v_{\rm crit}$ in cm s$^{-1}$ and $a_{\rm g}$ in cm), corresponding, respectively, to no coagulation or a critical velocity for coagulation which is consistent with the measurements of Poppe \& Blum (1997). The discussion so far has been concerned exclusively with the limit of no coagulation; now we shall compare with results obtained on including coagulation, with $v_{\rm crit} = 0.4$ $a_{\rm g}^{-\frac {5}{6}}$; see Fig.~\ref{coagulation}.

In Figs. \ref{coagulation}a and b are compared the results of calculations, with and without coagulation, of the fractional abundances of H$^+$, H$_3^+$ and its isotopes, and of the free electrons. From these Figures, it is apparent that the main effect of coagulation is that, at high densities, H$^+$ becomes the principal ion; this is because the rate of recombination of H$^+$ with electrons on the surfaces of (negatively charged) grains decreases with decreasing grain surface area and hence with the degree of coagulation (cf. Flower et al. 2005). To the higher density of H$^+$ there correspond lower values of the ortho:para ratios of H$_2$, H$_3^+$ and H$_2$D$^+$, as may be seen from Figs.~\ref{coagulation}c and d. In effect, the excited (ortho) states are more rapidly depopulated when the density of H$^+$ is higher, and so the ortho:para ratios decrease towards their values in Boltzmann equilibrium: $9\ {\rm exp}(-170.5/T)$ for the ortho:para H$_2$ ratio; $2\ {\rm exp}(-32.9/T)$ for the ortho:para H$_3^+$ ratio; and $9\ {\rm exp}(-86.4/T)$ for the ortho:para H$_2$D$^+$ ratio.

\section{Concluding remarks}

We have studied the population densities of molecules and molecular ions during the early stages of protostellar collapse, including H$_3^+$ and its deuterated isotopes, in their various nuclear spin states. The results of a free--fall simulation were compared with the corresponding steady--state calculation, for densities in the range $10^4 \le n_{\rm H} \le 10^7$~cm$^{-3}$. Calculations at this level of detail are required because current observations of the deuterated isotopes of H$_3^+$ (H$_2$D$^+$, D$_2$H$^+$) relate to {\it specific nuclear spin states}. The simulations show that there can be large discrepancies between the `free--fall' and `steady--state' results during the early stages of collapse.

In general, the computed population densities of the different nuclear spin states of molecules and molecular ions differ considerably from their values in Boltzmann equilibrium. A striking and important example is H$_2$, for which the ratio of population densities $n(J = 1)$ : $n(J = 0)$, computed in steady state for $n_{\rm H} = 10^4$~cm$^{-3}$ and $T = 10$~K, exceeds its value in Boltzmann equilibrium by 4 orders of magnitude. Furthermore, the value of the ortho:para H$_2$ ratio computed following the isothermal free--fall collapse exceeds the steady--state value by more than an order of magnitude at $n_{\rm H} = 10^5$~cm$^{-3}$. It is this overabundance which underlies the importance of ortho--H$_2$ in inducing population transfer between the different nuclear spin states of molecular ions, such H$_3^+$ and H$_2$D$^+$, at low temperatures.

In view of the long timescales associated with the processes involved, it seems unlikely that the ortho:para H$_2$ ratio will reach its steady--state value prior to the commencement of gravitational collapse. Present observations of the protostellar cores L1544 (Caselli et al. 2003) and IRAS 16293E (Vastel et al. 2004) are consistent with this statement; but the uncertainties in the observations and the limitations of the current model are such that definitive conclusions cannot be drawn. Our calculations show that values of the ortho:para H$_2$ ratio higher than in steady state lead to a reduction in the degree of deuteration of H$_3^+$. As H$_2$D$^+$ and D$_2$H$^+$ are the main agents of deuteration of other species, such as N$_2$H$^+$ and NH$_3$, a decrease in the level of deuteration of H$_3^+$ implies lower levels of deuteration of these other species. High values of the ortho:para H$_2$ ratio will occur earlier in the lifetime of the progenitor molecular cloud. Thus, one of the predictions of the present study is that protostars forming in `young' ($t \lesssim 10^6$~yr) molecular clouds  should {\it not} display high levels of deuteration, owing to the higher prevailing values of the ortho:para H$_2$ ratio.

\begin{acknowledgements}

GdesF and DRF gratefully acknowledge support from the `Alliance' programme, in 2004 and 2005. 

\end{acknowledgements}

\end{document}